\documentclass[aps,rmp,reprint]{revtex4-1}
\usepackage{graphicx}
\usepackage{color}
\usepackage{amsmath}
\usepackage{hyperref}

\newcommand{\citeasnoun}[1]{\citet{#1}}
\newcommand{\ack}[1]{\begin{acknowledgements}#1\end{acknowledgements}}

\begin{document}

\title{Nanoparticle characterization by continuous contrast variation in SAXS with a solvent density gradient}

\author{Raul Garcia-Diez}
\email[E-mail me at: ]{raul.garciadiez@ptb.de}
\author{Christian Gollwitzer}
\author{Michael Krumrey}
\affiliation{Physikalisch-Technische Bundesanstalt (PTB), Abbestr. 2-12, 10587 Berlin, Germany}

\begin{abstract}
Many low-density nanoparticles show a radial inner structure. This work proposes a novel approach to contrast variation with SAXS based on the constitution of a solvent density gradient in a glass capillary in order to resolve this internal morphology. Scattering curves of a polymeric core-shell colloid were recorded at different suspending medium contrasts at the four-crystal monochromator beamline of PTB at the synchrotron radiation facility BESSY II. The mean size and size distribution of the particles as well as an insight into the colloid electron density composition were determined using the position of the isoscattering points in the Fourier region of the scattering curves and by examining the Guinier region in detail. These results were corroborated with a model fit to the experimental data, which provided complementary information about the inner electron density distribution of the suspended nanoparticles.
\end{abstract}

\maketitle

\section{Introduction}

The morphology of nanoparticles determines the properties necessary for their utilization in real-world applications. For instance, in drug delivery devices the phenomena involved in biocompatibility reactions (e.g. protein adsorption) depend on the amount of available surface and the nanoparticles' properties \cite{Vittaz1996}. Particularly, polymer lattices and biodegradable nanoparticles have been of growing importance of late as drug carriers \cite{Kattan1992} and thus extensively characterized \cite{Soppimath2001}. The size determination and the characterization of the radial structure of the particles are therefore fundamental tasks. 

The contrast variation method in Small Angle X-ray Scattering (SAXS) experiments consists in systematically varying the electron density of the dispersing media to study the different contributions to the scattering intensity in greater detail as compared to measurements at a single contrast. It emerges as an ideally suited technique to elucidate the structure of particles with a complicated inner composition and has been repeatedly employed to investigate the radial structure of latex particles suspended in an aqueous medium \cite{Dingenouts1999,Ballauff2011}. In Small Angle Neutron Scattering (SANS) the contrast variation technique is widely used by mixing water and deuterium oxide, but the use of deuterated chemicals and the incoherent contribution to the background as well as the limited access to neutrons restrict the application of this technique. Other methods for structural investigation (e.g. transmission electron microscopy \cite{Joensson1991,Silverstein1989}) require prior treatment of the sample and are not ensemble averaged. 

In SAXS, the solvent contrast variation technique is achieved by adding a suitable contrast agent to the suspending medium (e.g. sucrose) and recording the scattering data as a function of the adjusted solvent electron density \( \rho_{solv} \) \cite{Ballauff2001,Bolze2003}. In order to resolve small changes of the radial structure, the average electron density of the colloidal particles must be close to the dispersant's, i.e., the \emph{match point} should be approached, where the average contrast of the particle vanishes. In the case of polymeric lattices with electron densities ranging from 335 to \(390 \mbox{ nm}^{-3}\), an aqueous sucrose solution is very well suited as the suspension medium, due to the easy realization of concentrated solutions with electron densities of up to \(400 \mbox{ nm}^{-3}\). Previous studies on globular solutes \cite{Kawaguchi1992} and the influence of the sucrose on the size distribution of vesicles \cite{Kiselev2001} show the feasibility of this technique, while further studies have investigated the effect of the penetration of the solvent into the particles \cite{Kawaguchi1993}.

The preparation of a number of different sucrose solutions has been a major inconvenience in solvent contrast variation experiments, due to the tedious, time-consuming process, possible inaccuracy in the sucrose concentration and the discrete range of available solvent electron densities. In this article we propose a novel approach using a density gradient column, which allows the tuning of the solvent contrast within the provided density range, resulting in a virtually continuous solvent contrast variation. By filling the bottom part of the capillary with a particle dispersion in a concentrated sucrose solution and the top part with an aqueous solution of the same particle concentration, a solvent density gradient is initiated with a constant concentration of nanoparticles along the capillary. Density gradient columns are extensively used in fields like marine biology \cite{Coombs1981} or biochemistry together with centrifugation \cite{Hinton1978}, to create a continuously graded aqueous sucrose solution by diffusion of the sucrose molecules. Combining this approach with SAXS, it is possible to choose \emph{in situ} the most appropriate solvent densities to perform measurements close to the contrast match point and to acquire extensive datasets in a short interval of time through the high brilliance and collimation of current synchrotron radiation sources. These datasets can be analysed using different, complementary evaluation methods. In this article, both a model-free theoretical framework as well as model fit are applied and, in combination, deliver a detailed insight into the inner structure of particles. 

In order to demonstrate the proposed technique, latex nanoparticles with a core-shell structure were measured. The particles have a narrow size distribution and consist of a spherical polystyrene (PS) core enclosed by a thin shell of a denser polymer, most likely poly(methyl methacrylate) (PMMA). This is presented according to the following structure. Firstly in \(\S\)\ref{sec:theory}, the underlying theory of SAXS contrast variation is briefly reviewed and the scattering form factor used for the model fitting is presented. The details of the experimental data acquisition are shown in \(\S\)\ref{sec:experiment}, followed by SAXS data evaluation using different methods in \(\S\)\ref{sec:results}, jointly with a discussion of the experimental measurements and a consistency check of the obtained results. Finally, in \(\S\)\ref{sec:conclusion} the experimental results of the particle size distribution and radial structure are summarized and the applicability of the solvent contrast variation technique in SAXS is discussed.

\section{Theoretical background and contrast variation} 

\label{sec:theory}
The scattering intensity of an ensemble of randomly oriented nanoparticles in suspension can be expressed as a function of the momentum transfer \( q \), modulus of the scattering vector \(\vec q\), as
\begin{equation}
\label{eq:intensity}
I(q)=N\int_{0}^{\infty} g(R)\left|F(q,R) \right|^2 dR,
\end{equation}
where \(N\) is the number of scatterers, \(g(R)\) is the size distribution function and \(F(R)\) is the particle form factor, which depends on the inner radial structure of the particle. If the particle shows an heterogeneous morphology, the form factor differs qualitatively for different suspending medium densities.  For sufficiently monodisperse particle suspensions, the Fourier region of the scattering curve shows pronounced minima that characterize the particle structure. The next sections describe different possibilities to evaluate the scattering data.

\subsection{Form factor analysis}
When analyzing contrast variation data, a widespread theoretical approach is based in the non-interacting model proposed by Stuhrmann $\&$ Kirste (\citeyear{Stuhrmann1965,Stuhrmann1967}) for monodisperse particles. The so-called \emph{basic functions} formulation differentiates, independently of the particle inner structure, the contributions which depend on the varying solvent density or contrast (\(\Delta\eta=\rho_{core}-\rho_{solv}\)) and on the excess of electron density of each component \(\Delta \rho_i =\rho_i-\rho_{core}\). 

For a typical morphology with sharp interfaces between the radial symmetric components of the particle with radius \(R_i\) the form factor is
\begin{equation}
\label{eq:multicore-shell}
	\begin{split}
		F\left(q,R \right)= & \Delta \eta f_{sph}(q,R)+\\
		&\sum_{i=1}^{n-1} \Delta\rho_i \left( f_{sph}(q,R_{i+1})-f_{sph}(q,R_{i}) \right) ,
	\end{split}
\end{equation}
where \(R\) is the external radius of the particle, \( n \) is the number of concentric shells and \(f_{sph}\) is the form factor of a homogeneous solid sphere given by
\begin{equation}
f_{sph}(q,R)=\frac{4}{3} \pi R^3  \left( 3\frac{\sin{qR}-qR\sin{qR}}{(qR)^3}\right).
\label{eq:ff_sph}
\end{equation}

\subsection{Isoscattering point}
One of the best known features appearing in a contrast variation experiment is the existence of \emph{isoscattering points}. At these specific \( q\)-values, the scattering intensity is independent of the adjusted solvent contrast, i.e. all scattering curves intersect in the isoscattering points regardless of the contrast. The isoscattering points \(q^{\star}\) are particularly interesting because they emerge for any spherical particle with an inner structure and a sufficiently narrow size distribution. From the contrast-depending part of equation \eqref{eq:multicore-shell}, a model-free expression can be derived which relates the position of the isoscattering points \(q^{\star}_i\) with the external radius of the particle \( R \), independent of its radial structure \cite{Kawaguchi1992}:
\begin{equation}
\label{eq:isoscattering}
\tan(q^{\star}_iR)=q^{\star}_iR
\end{equation}
The positions of the isoscattering points correspond to the minima positions of the scattering intensity of a compact spherical particle with radius \( R \). Although this expression is derived for the monodisperse case, it can still be applied up to a moderate degree of polydispersity, if care is taken regarding the shift of the minima position due to polydispersity \cite{Beurten1981}. If defining the polydispersity degree \( p_d\) as the full width half maximum of the particle size distribution divided by its average value, for size distributions with \( p_d\) larger than \( \approx 30\,\% \), the isoscattering point is not well defined and the intersection point of the curves is smeared out, showing a diffuseness in the isoscattering point position \cite{Kawaguchi1992}.

\subsection{Radius of gyration and intensity at zero angle}
\label{sec:guinier}
The radius of gyration \( R_g\) is systematically employed in small-angle scattering as an evaluation tool \cite{Mertens2010,Sim2012}. It can be calculated using the Guinier approximation \cite{Guinier1939,Guinier1955}, which assumes that the scattering intensity behaves in the limit of small \(q\) as
\begin{equation}
\label{eq:guinier}
I(q)=I(0)\,\mbox{exp}\left(-\frac{R_g^2}{3}q^2\right),
\end{equation}
where \( I(0)\) is known as forward scattering or intensity at zero angle. Using the basic functions approach, the radius of gyration of a monodisperse, heterogeneous particle can be expressed as a function of the solvent electron density \( \rho_{solv} \) and the average electron density of the particle \( \rho_0 \) \cite{Feigin1987}
\begin{equation}
R_g^2=R_{g,c}^{\,2}+\frac{\alpha}{\rho_0-\rho_{solv}}-\frac{\beta}{(\rho_0-\rho_{solv})^2},
\label{eq:gyration}
\end{equation}
where \(R_{g,c}\) is the radius of gyration of the particle shape corresponding to the volume inaccessible for the solvent \( V_c \), \( \alpha \) characterizes the distribution of different phases inside the particle and \( \beta>0 \) considers the eccentricity of the different scattering contributions \cite{Stuhrmann2008}. Nevertheless, particle aggregation influences the scattering curves especially in the Guinier region and must be explicitly avoided.

\citeasnoun{Avdeev2007} proposed an extended version to equation \eqref{eq:gyration} for the case of a polydisperse particle ensemble by introducing the \emph{effective} values \( \tilde R^2_{g,c} \), \( \tilde \alpha \) and \( \tilde \beta \), which are the intensity-weighted averages of the corresponding parameters over the polydispersity. The observed average electron density is not affected by the polydispersity (\( \tilde\rho_0=\rho_0 \)) if the volume ratio between the different particle components is constant for all particles in the ensemble.

Assuming the same premise, the intensity at zero angle is given by
\begin{equation}
\label{eq:I0}
I(0)\propto N \left( \rho_0-\rho_{solv} \right)^2 ,
\end{equation}
with a minimum at \( \rho_{solv}=\rho_0 \). Hence, by analyzing the Guinier region of the scattering curves, it can be obtained the average electron density of the particle without assuming an \emph{a priori} inner structure.

Using the previously models presented above, it is possible to obtain by independent means the external radius and the average electron density of the particle in suspension.

\section{Experimental methods}
\label{sec:experiment}

The preparation of a polymeric nanoparticle suspension density gradient within a glass capillary using an aqueous sucrose solution, the X-ray transmittance measurements at different positions along its vertical axis and the collection of scattering patterns at the calibrated capillary positions with distinct contrasts are described in the following sections. 

\subsection{Solvent density gradient preparation}
\label{sec:gradient}
Carboxylated polystyrene nanoparticles with a nominal size of 105 nm suspended in water were purchased from Kisker Biotech (Steinfurt, \emph{Germany}). The synthesis by multi-addition emulsion polymerization suggests that the assumption made in \(\S\)\ref{sec:guinier} is correct and the average density of the particle is not altered by the size polydispersity.

The solvent density gradient was prepared in vacuum-proof borosilicate glass capillaries from Hilgenberg (Malsfeld, \emph{Germany}) with a rectangular cross section of \( (4.2\pm0.2)\times(1.25\pm0.05)\mbox{ mm}^2\),  a length of \((80\pm0.5) \mbox{ mm} \) and a wall thickness of ca. \(120\;\mu\mbox{m}\). The bottom end of the capillary was closed by welding and the lower section, up to a height of ca. \(1\) cm, was filled with Galden\textregistered PFPE SV90 from Solvay Plastics (Brussels, \emph{Belgium}). This fluid has an exceptionally high density of $1.69\,\mathrm{g/cm^3}$, low viscosity and is immiscible with aqueous solutions. Consequently, a uniform interface with the particle suspension is formed at the bottom. 

Directly above the Galden fluid, the denser of two mixtures with different solvent densities and an equal particle concentration of \(12.6 \mbox{ mg/ml} \) was filled into the capillary using a syringe up to a height of  \(9\) mm. The dense aqueous solution was prepared with \( 21.23\,\%\) sucrose mass fraction (Sigma-Aldrich (Missouri, \emph{USA})) with a physical density of \(\rho_1=1.088 \) g/cm\(^3\), whereas a lighter one was produced without sucrose (\(\rho_2=0.997 \) g/cm\(^3\)). The light mixture was then filled on top of the aqueous sucrose solution along ca. \(8\) mm. By the time the two components come into contact, the density gradient is started with density values \(\rho_1\) and \(\rho_2\), a total gradient length \(L=17\) mm and the interface position at \(z_0=9\) mm. The calculated diffusion timescale of the solvent density gradient is ca. 10 minutes, considering the diffusion coefficient \(D=5.2 \cdot 10^{-10} \;\frac{\mbox{m}^2}{\mbox{s}}\) \cite{Uedaira1985,Ribeiro2006} and assuming that convection effects are negligible due to the small length-scale of the capillary \cite{Berberan-Santos1997}. The time needed for the transfer of the sample into the high vacuum chamber amounts to ca. 1 hour. Within this time duration, the deviation of the solvent density at both ends of the gradient from the initial value can be estimated with an uncertainty below \(0.5\,\%\). If the same capillary is measured at different points in time during the diffusion process of the sucrose, several data sets with different solvent densities can be recorded and a very dense data set with a virtually continuous variation in the suspending medium density can be achieved.

\subsection{Experimental setup: X-ray measurements}

The measurements were performed at the four-crystal monochromator beamline in the PTB laboratory at the electron storage ring BESSY II (Berlin, \emph{Germany}), which provides highly intense, collimated synchrotron radiation focused on the sample and collimated into a \(0.5\) mm circular spot by Ge pinholes situated between the sample and the monochromator with an energy resolving
power \( E/\Delta E \) of \( 10^4 \). To measure the total flux and sample transmission, photodiodes were used which were calibrated against a cryogenic electric substitution radiometer with a relative uncertainty of \( 1\,\% \) \cite{Krumrey2001}.

\begin{figure}
\caption{The rectangular density gradient capillary is placed in the X-ray beam and can be moved by sample motors in both directions perpendicular to the incoming beam.}
\scalebox{0.95}{\includegraphics{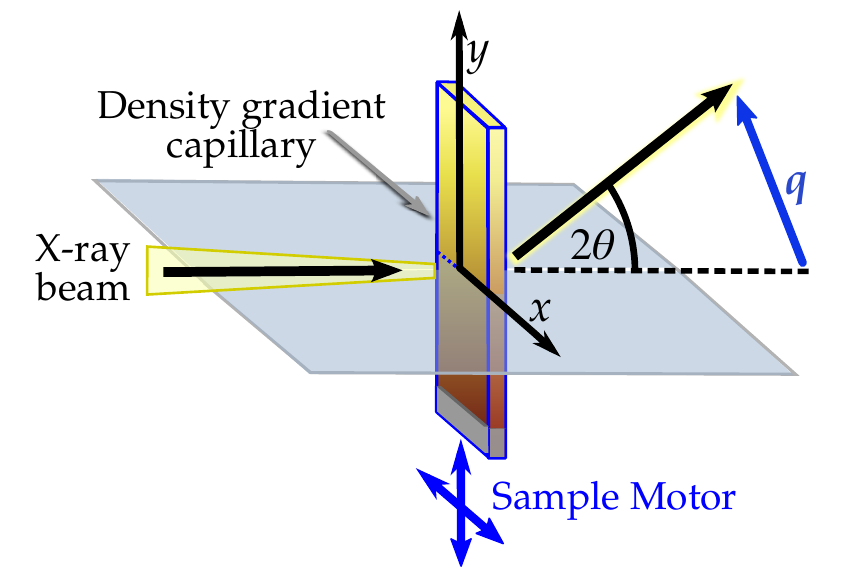}}
\label{fig:0}
\end{figure}

The rectangular capillary is placed in a sample holder which allows the movement with micrometer precision in the directions perpendicular to the incoming beam, as depicted in figure \ref{fig:0}. In order to determine the central vertical capillary axis, a horizontal X-ray transmission scan is performed at two different vertical positions of the capillary spaced by 20 mm. The central vertical axis can be drawn from the centers of both measurements and the sample can be moved along this axis by the simultaneous operation of the vertical and horizontal motors.

\subsection{X-ray transmission}
\label{sec:transmission}
\begin{figure}
\caption{Calculated aqueous sucrose solution density along the gradient capillary vertical axis from the transmission measurements. The dotted line displays the measurement at a diffusion time \(t_1=78\mbox{ min}\), the dashed line shows the measurement after recording the scattering patterns (15 minutes later) and the thick line shows the calibrated value used.}
\scalebox{0.45}{\includegraphics{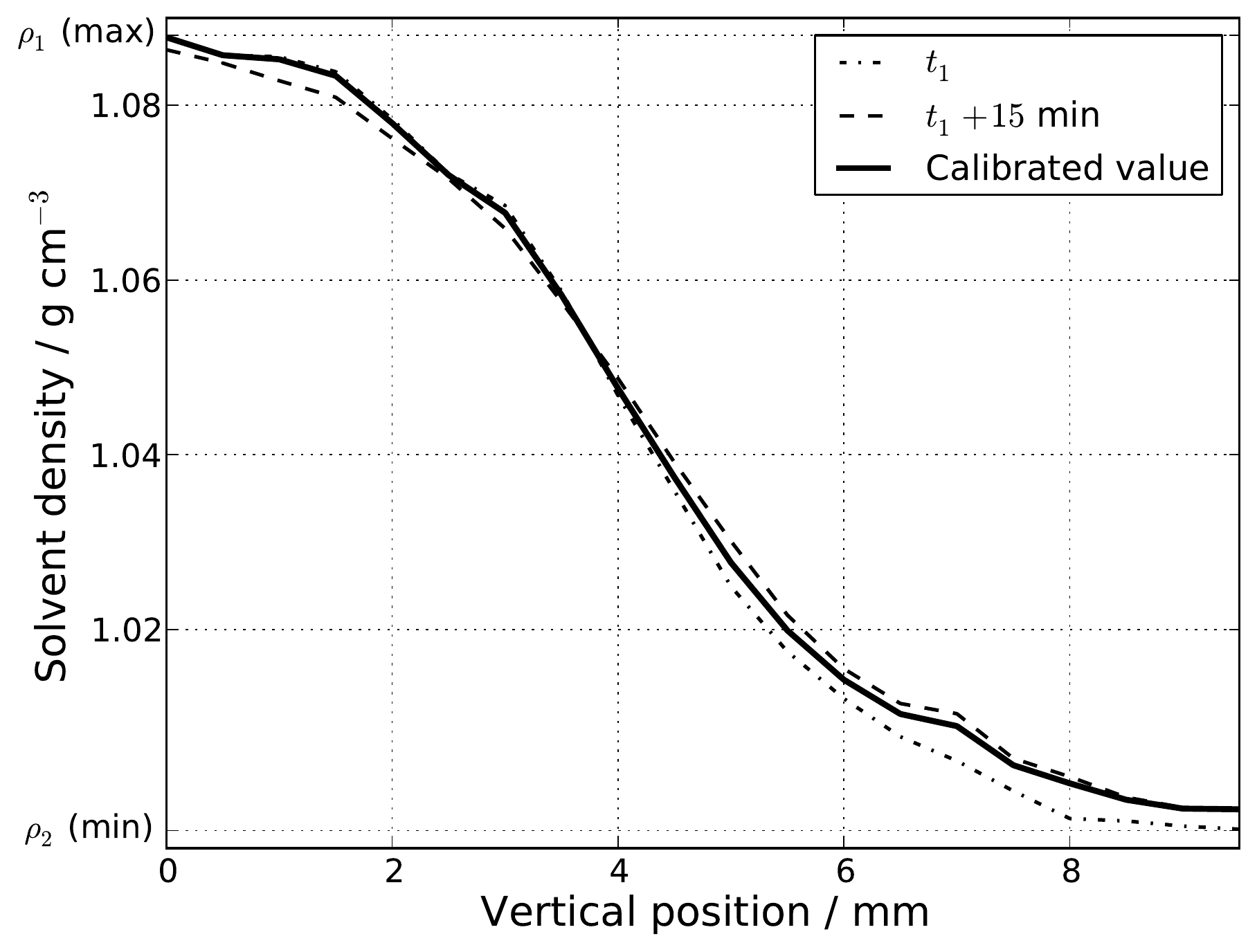}}
\label{fig:1}
\end{figure}

The transmitted intensity through the sample is recorded at a photon energy of \(E = (5500.0 \pm  0.5)\) eV for 10 seconds at each position. The measurement consists of 20 points spaced 0.5 mm along the central vertical axis of the capillary. The overall X-ray transmission measurement requires approximately 5 minutes, which is within the calculated diffusion timescale of the aqueous sucrose solution. The solvent electron density profile within the density gradient capillary derived from this measurement is depicted in figure \ref{fig:1}. A uniform thickness of the capillary within $0.5\,\%$ along this axis was determined by measuring the X-ray transmission of an empty capillary. The associated uncertainty in the sample transmission measurement is below $4\,\%$. The sample thickness is assumed to be constant. This transmission measurement is performed both immediately before and after recording the scattering patterns, which takes 15 minutes to complete. The transmittance values used for the density calibration are then linearly interpolated between both data sets taking into account the time-dependence. These values can be converted to solvent electron densities via the Beer-Lambert law, which relates the density of the solution with the transmitted intensity:
\begin{equation}
  \rho(z) = A \left( \ln{I_0} - \ln{I(z)} \right) .
\end{equation}
Here \(\rho\) is the electron density of the suspending medium, $I$ and $I_0$ are the transmitted and incoming intensities respectively and $A$ is a factor determined by the reference values of the solvent electron density at the vertical limits of the capillary at the initial time. The sucrose concentration in solution expressed as the mass fraction \( M \) at these reference points can be converted to electron densities with the empirical formula \( \rho=1.2681M+333.19 \) nm\(^{-3}\) \cite{Haynes2012}. The suspending medium electron density shows a maximum uncertainty of 1 nm$^{-3}$ associated with the vertical size of the focused X-ray beam.

\subsection{SAXS measurements}
\label{sec:SAXS}
The sample was moved in steps of 0.5 mm along the central vertical capillary axis and exposed at each position for 45 seconds. At these positions, the solution transmittances were previously measured and the suspending medium electron density calibrated. The measured scattering curve is an average over a range of solvent electron densities associated with the beam size. The momentum transfer \(q\) of the scattering curves was calculated using
\begin{equation}
q=\frac{4\pi E}{hc}\sin\theta ,
\end{equation}
where \(\theta\) is half of the scattering angle, \(h\) is the Planck constant and \(c\) is the speed of light. The incident photon energy \(E = \left(8800.0  \pm 0.8\right)\) eV was chosen to be higher than the photon energy for the transmission measurements to improve the recorded statistics, due to a ca. \(150\) higher transmission \cite{Henke1993}. The scattered X-ray photons were collected with a vacuum-compatible Pilatus 1M hybrid-pixel detector (Dectris Ltd, (Baden, \emph{Switzerland})) with a pixel size of \(d = \left(172.1  \pm 0.2\right) \; \mu \)m at a distance \(L = \left(4540.2  \pm 0.8\right)\) mm from the capillaries, determined by triangulation using a calibrated length measurement system \cite{Wernecke2014}. The obtained scattering curve was normalized to the exposure time and the incident intensity, measured by means of a calibrated transparent silicon diode.  In total, 40 scattering curves with different solvent electron densities were measured at two different times \(t_1=\)78 min and \(t_2=\)156 min after filling the capillaries.

\section{Results and discussion}
\label{sec:results}
\begin{figure}
\caption{Experimental scattering curves of polystyrene nanoparticles for different suspending medium electron densities measured between 78 and 93 minutes after the inception of the density gradient. The dashed line shows the experimental background, containing scattering contributions from the capillary and the pure solvent.}
\scalebox{0.29}{\includegraphics{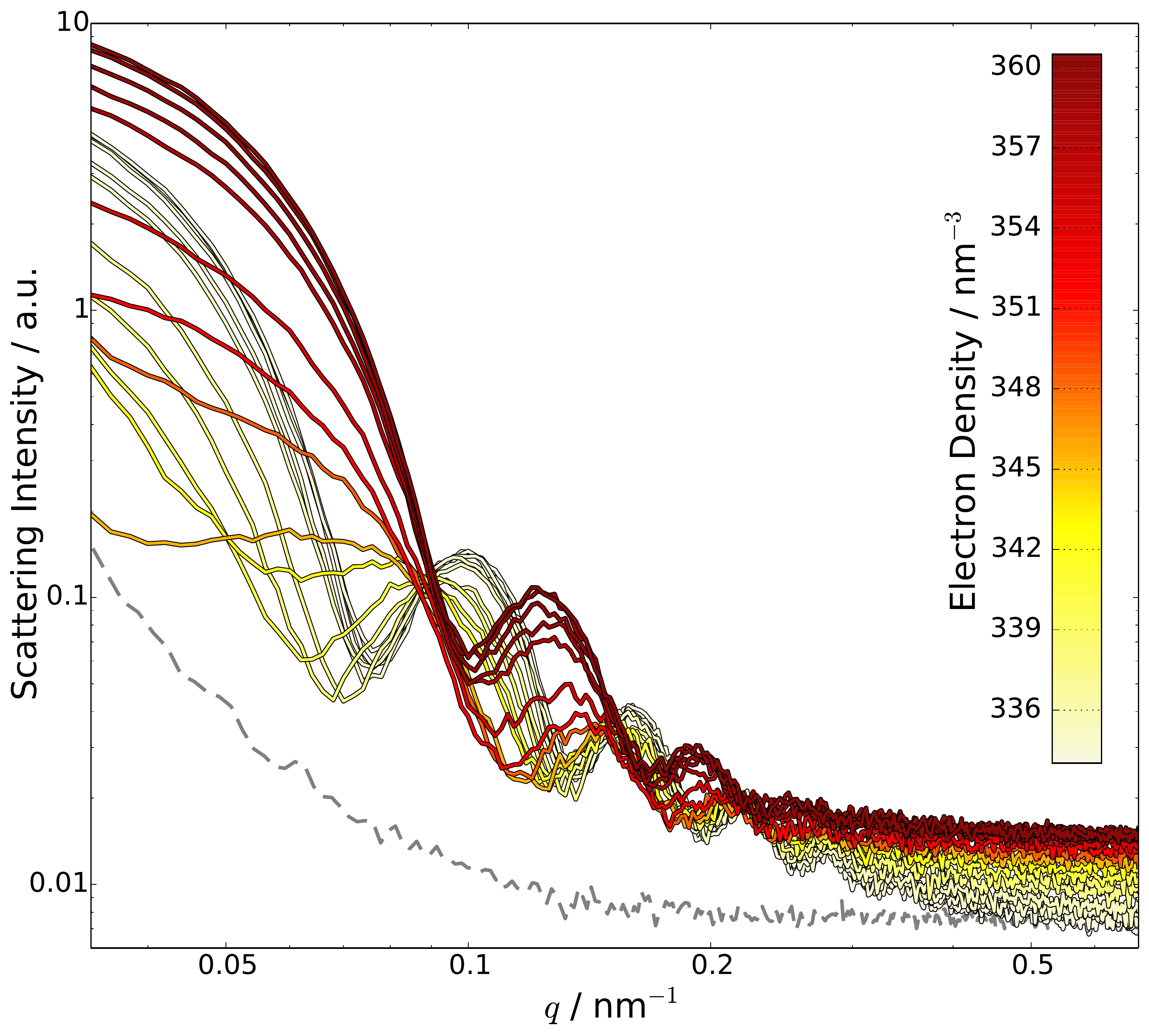}}
\label{fig:2}
\end{figure}

The measured scattering curves of the polystyrene particles are displayed in figure \ref{fig:2}. In the region for \(q\) from 0.03 nm\(^{-1}\) to 0.5 nm\(^{-1}\) it is possible to observe the variation of the curve features corresponding to the particle form factor through the increase of the solvent electron density from 333.7 nm\(^{-3}\) at the top edge of the density gradient to 360.3 nm\(^{-3}\) at the maximum sucrose concentration. In this region, the experimental background is composed mainly by the contribution of the capillary scattering at the low $q$-region and the uniform scattering of the suspending medium. The scattering background intensity lies one order of magnitude below the sample scattering curves along most of the available $q$-range. Furthermore the solvent background barely differs for the different aqueous sucrose concentrations and its differences have a small contribution to the scattering curve compared to the features in the Fourier region.

\begin{figure}
\caption{The thick red line shows the scattering curve measured at $\rho_{solv}=345.4$ nm$^{-3}$, close to the match point, and the dotted line displays the experimental background. The symbols with errorbars show the background corrected scattering curve.}
\scalebox{0.3}{\includegraphics{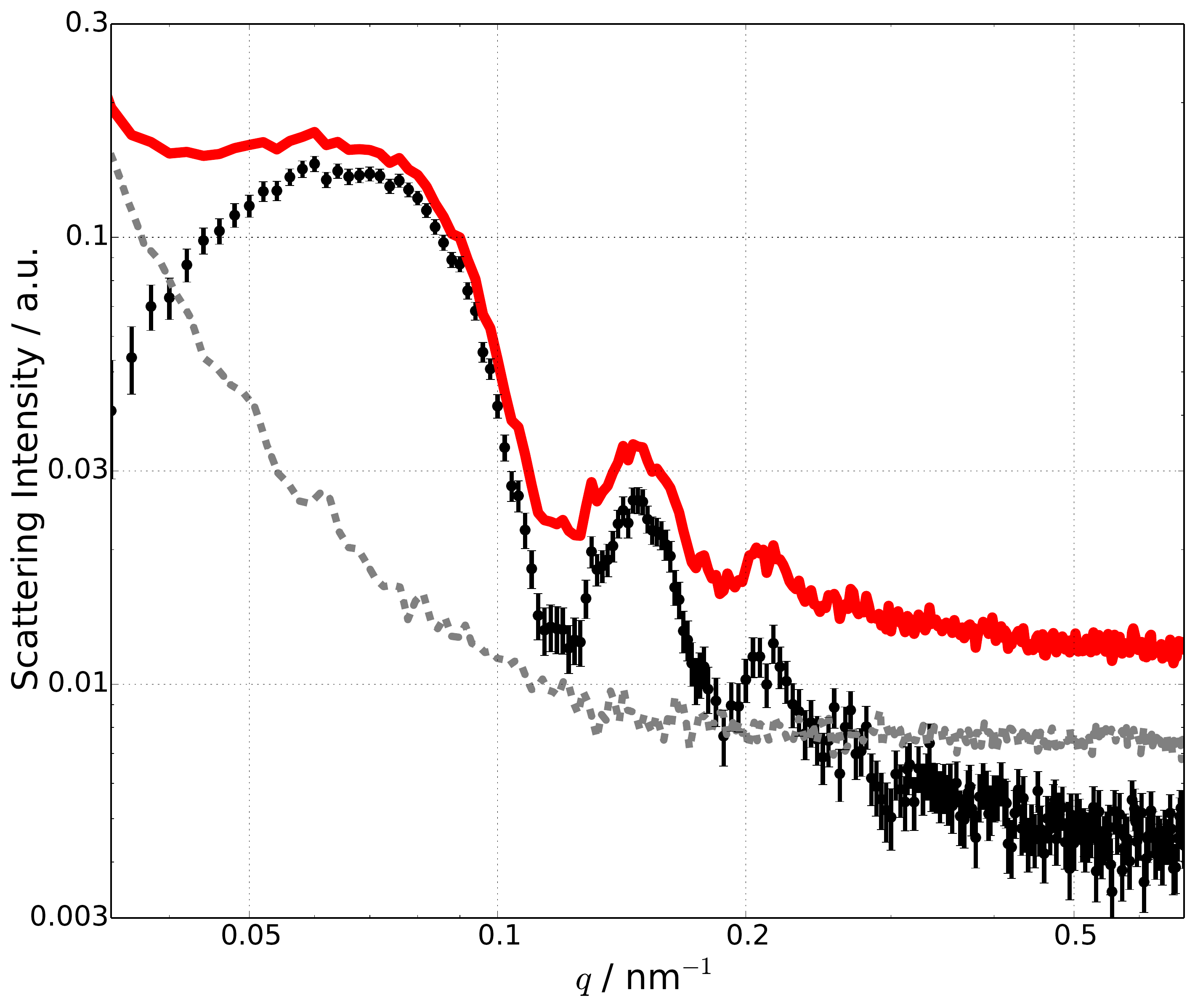}}
\label{fig:2B}
\end{figure}

Upon increasing the solvent density, the position of the first minimum shifts from 0.07 nm\(^{-1}\) towards smaller \(q\)-values until it vanishes when the solvent electron density matches the average electron density of the measured particle. In the Fourier region of the scattering curves, several minima are observed which shift towards smaller \(q\)-values when increasing the solvent electron density. Upon subtracting the experimental background from the scattering curve, a decrease of the scattering intensity towards $q=0$ is observed only for the solvent electron density closest to the match point as depicted in figure \ref{fig:2B}. Therefore, background corrections for the whole dataset can be neglected. The behaviour at low $q$-values will be further discussed in section \S\ref{sec:guinier_analysis} when evaluating the zero-angle intensity.

The presence of the clearly visible isoscattering point around \(q=0.09\) nm\(^{-1}\) confirms the existence of an inner structure. This heterogeneous composition was previously reported for the same colloids by \citeasnoun{Minelli2014}, who observed methacrylic acid (MAA) and methylmethacrylate (MMA) at the particle surface, both monomer precursors of PMMA polymerization. A more detailed insight into the radial morphology is presented subsequently, using the theoretical framework already introduced.

\subsection{Core-shell model fit}
\label{sec:model_fit}
\begin{figure}
\caption{The simulated scattering curves from the core-shell model fit at three selected contrasts \( \rho_0-\rho_{solv} \) are shown as lines together with the experimental data points. In the inset, the electron density profile corresponding to the fitted core-shell form factor is displayed.}
\scalebox{0.45}{\includegraphics{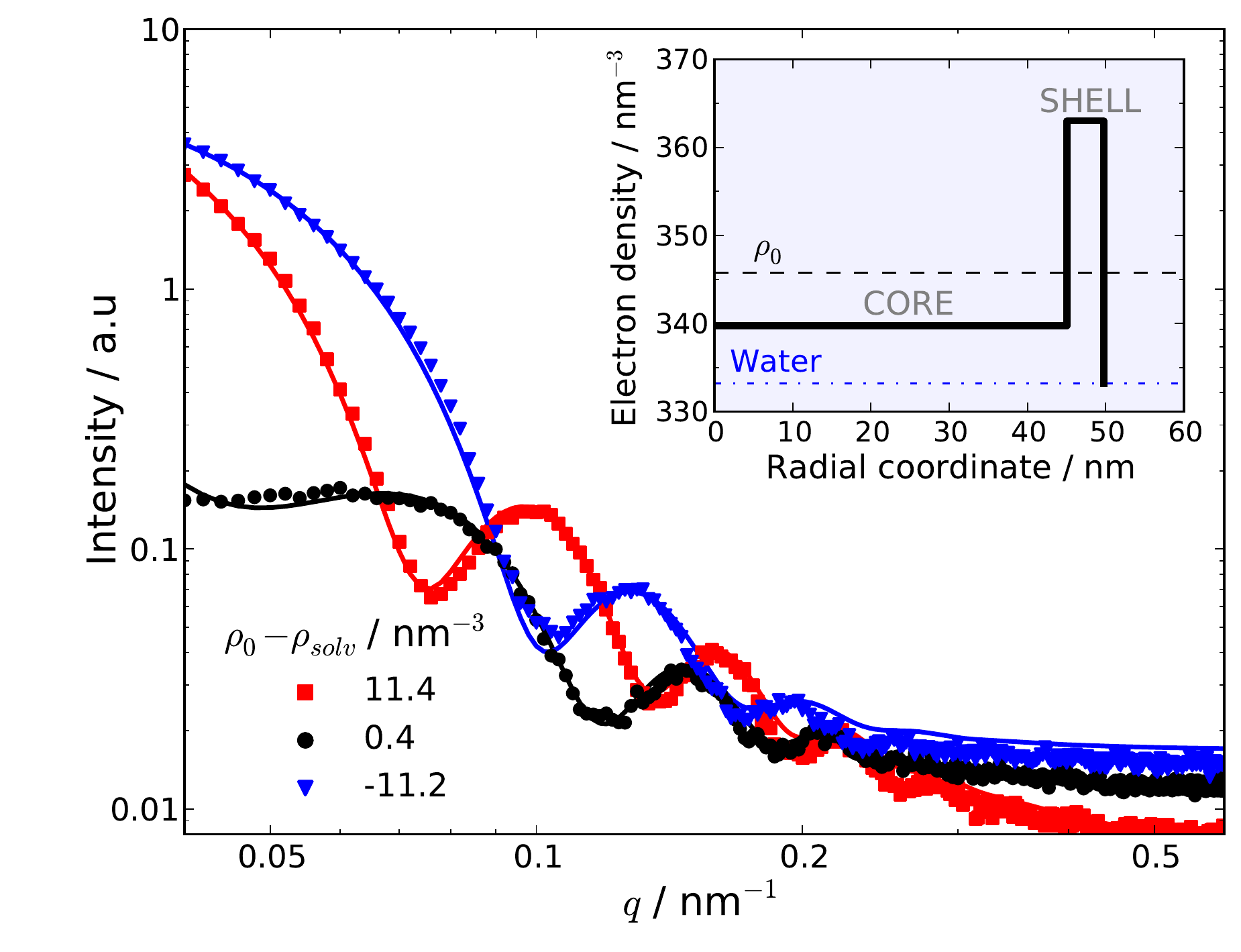}}
\label{fig:3}
\end{figure}

A core-shell model fit to the scattering curves is displayed in figure \ref{fig:3} for three representative contrasts. The model represents a radially symmetric particle, with a sharp interface between the outer shell and the inner core. This is a specific case of equation \eqref{eq:multicore-shell} with \( n=2 \)
\begin{equation}
	\begin{split}
	F_{CS}(q,R,R_{core})= & \Delta\eta f_{sph}(q,R)+ \\
	& \Delta\rho\left[ f_{sph}(q,R)-f_{sph}(q,R_{core}) \right] ,
	\end{split}
\label{eq:ff_cs}
\end{equation}
where \(R \) and \(R_{core} \)  are the outer shell and inner core radii respectively and the excess of electron density is \(\Delta\rho=\rho_{shell}-\rho_{core}\). The simultaneous fitting of the form factor to the 40 measured scattering curves was performed by means of the method of least squares in the Fourier region \cite{Pedersen1997}. The calculated scattered intensity was modelled as the sum of a constant and a power-law background \(I_{bg}=C_0+C_4q^{-\alpha} \) and the particle contributions. The fitted parameters were \(\rho_{core}\), \(\rho_{shell}\), \(R\), \(R_{core}\) and \(\alpha\), whilst a Gaussian size distribution was assumed. For the suspending medium electron density \( \rho_{solv} \) appearing in the contrast \( \Delta\eta \), the value determined from the transmission measurement was used for each curve.
The obtained results are \(R=\left(49.7 \pm 2.8\right) \) nm, \(R_{core}=\left(44.2 \pm 0.9\right) \) nm, \(\rho_{core}=\left(339.7 \pm 0.1\right)\) nm\(^{-3}\) and \(\rho_{shell}=\left(361.9 \pm 2.0\right)\) nm\(^{-3}\), which represent the radial structure of a dense, thin shell surrounding a lighter core, as seen in the inset of figure \ref{fig:3}. The resulting average electron density of the particle is \(\rho_{0}=\left(345.9 \pm 1.5\right)\) nm\(^{-3}\) and the polydispersity degree, \(p_d=\left(22.8\pm 6.0\right)\,\%\). The best fitting background corresponds to a value of \( \alpha = 4.3\pm 0.5 \), close to the case \( \alpha = 4 \) originating from large impurities or precipitates \cite{Pedersen1994}. The fit uncertainty was calculated with a confidence interval of one standard deviation. 

It is noticeable that the calculated electron density of the core coincides exactly with the theoretical polystyrene electron density, although the electron density of the shell is remarkably lower than the theoretical value of 383.4 nm\(^{-3}\) for PMMA \cite{Ballauff2001}. This might arise from the lower density of the monomers used in the particle synthesis (MAA and MMA), which could have mixed with the styrene monomers resulting in a less dense material than PMMA. This model might present some differences with the real colloid system, as a diffusive interfacial layer can be expected between polymer phases in colloids \cite{Dingenouts1994}, especially for incompatible polymers such as PMMA and PS. On the other hand, the large quantity of scattering curves used for the fitting process and, accordingly, the decreased uncertainty suggests that the chosen sharp core-shell model has a great resemblance to the real particle.

\subsection{Isoscattering point}
\begin{figure}
\caption{Relative standard deviation of the scattering curves as a function of the momentum transfer. The labelled minima correspond to the first five isoscattering point positions calculated by fitting a Lorentzian function (red line).}
\scalebox{0.45}{\includegraphics{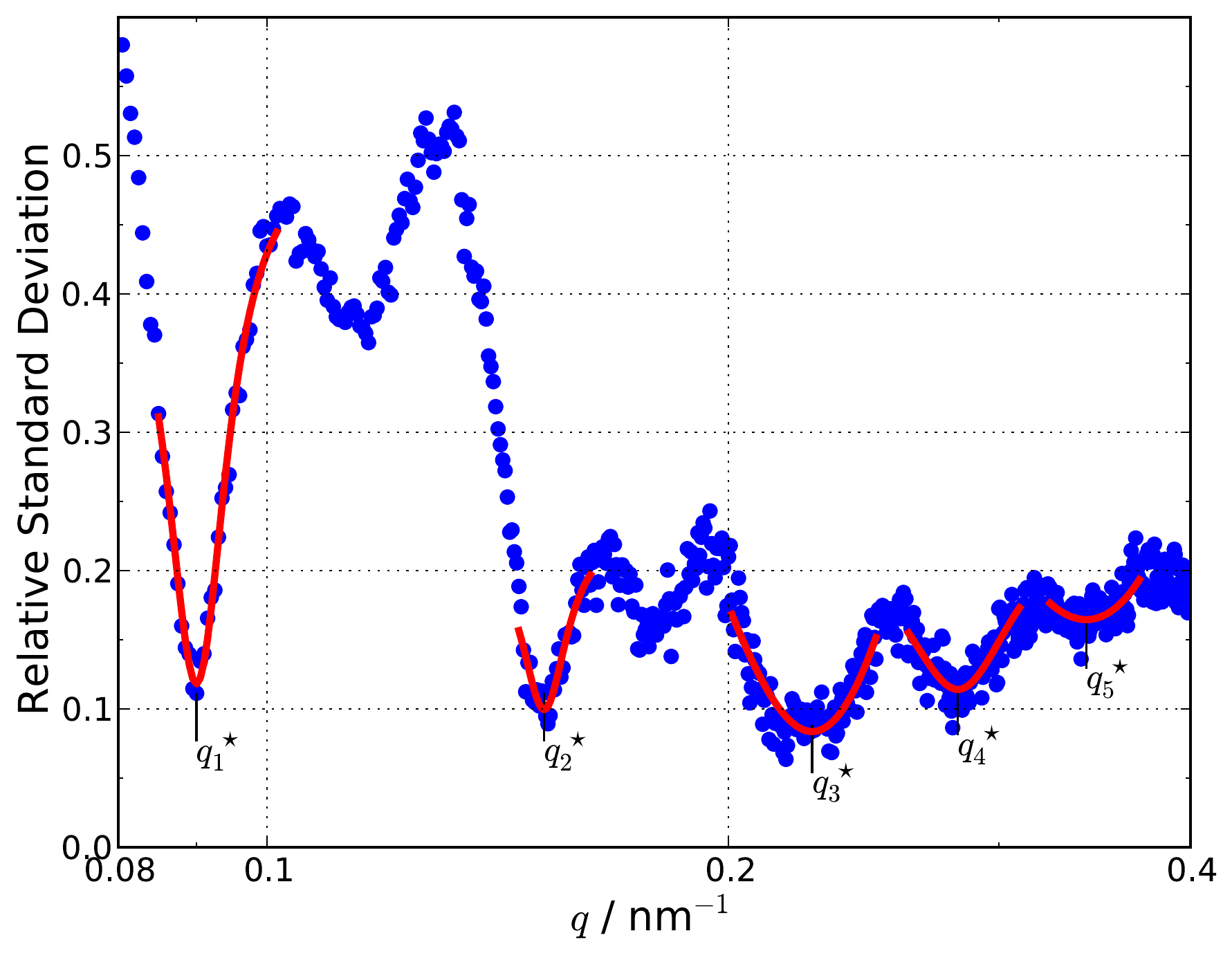}}
\label{fig:4}
\end{figure}

The first isoscattering point is clearly visible in figure \ref{fig:2}. For a more quantitative evaluation, the relative standard deviation of the 40 measured curves at each \(q\) is calculated according to
\begin{equation}
\sigma_r (q)=\frac{1}{\bar{I}(q)}\sqrt{\frac{\sum^{M}_{i=1} (I_i(q) -\bar{I} (q))^2 }{M-1}} ,
\end{equation}
where \(\bar{I} (q)\) is the mean value of the intensity at \(q\) and \( M \) is the number of scattering curves. This value becomes minimal at an isoscattering point. In order to reduce the influence of outliers, a truncated mean value was utilized, disregarding the \(10\,\%\) most dispersed data points. In figure \ref{fig:4}, the relative standard deviation is plotted as a function of the momentum transfer \(q\), which shows several distinguishable minima corresponding to isoscattering points.

A precise determination of the isoscattering point positions is performed by fitting Lorentzian functions to the minima in the relative standard deviation plot, which allows the calculation of the model-free external radius of the particle by means of equation \eqref{eq:isoscattering}. The results are presented in table \ref{tab:polydispersity}. The obtained particle radii vary in the range from \(48.1\) nm to \(51.0\) nm, although as predicted by \citeasnoun{Kawaguchi1992} for a polydisperse system, the isoscattering points get smeared out for larger \( q \)-values and the precision decreases, simultaneously with the increase of the solvent background at higher \(q\)-values. This can be directly observed in the quality of the experimental data, as the first two minima are clearly pronounced, while the subsequent minima appear smeared out. For instance, the isoscattering point \(q^{\star}_5\) is already too weak for an accurate evaluation and the third minimum shows two remarkably close smaller minima which might affect the shape of the function. Therefore, \(q^{\star}_1\) and \(q^{\star}_2\) yield the most reliable values for evaluating the external radius of the particles, although all results are presented in table \ref{tab:polydispersity}. The value derived from the isoscattering points \(R=50.5\) nm differs by only \(1.6\,\%\) from the radius calculated from the model fit in the previous section.

\begin{table}
\caption{Experimentally determined position of the first five isoscattering points and the corresponding external particle radius \(R\).}
\begin{tabular}{l|cc}
 & \( q^{\star} \) (nm\(^{-1}\))    &  \(R\) (nm) \\
\hline
 \(q^{\star}_1\) &  0.0900 & 49.9 \\
 \(q^{\star}_2\) &  0.1516 & 51.0  \\
 \(q^{\star}_3\) &  0.2267 & 48.1   \\
 \(q^{\star}_4\) &  0.2822 & 49.9    \\
 \(q^{\star}_5\) &  0.3421 & 50.3     \\
\end{tabular}
\label{tab:polydispersity}
\end{table}

Due to the ambiguous definition of the isoscattering point diffuseness, a quantitative determination of the polydispersity of the suspended nanoparticles by means of the Lorentzian profile is rather challenging. Nevertheless, the narrow size distribution of the sample becomes clear by comparing the relative standard deviation values of the observed minima in figure \ref{fig:4} with a simulation using the structural parameters obtained in section \S\ref{sec:model_fit}. The value \( \sigma_r(q^{\star}_1)=0.11 \) corresponds to a calculated ensemble polydispersity of \(24\,\% \). This value serves as an upper \( p_d \) limit due to the possible overestimation caused by the scattering contribution of the suspending medium.

\subsection{Guinier region analysis}
\label{sec:guinier_analysis}
By analyzing the low \(q \)-region of the scattering curves, the so-called Guinier region, two important parameters can be obtained: the radius of gyration \(R_g\) and the intensity at zero angle \(I(0)\). According to \citeasnoun{Feigin1987}, the fit of equation \eqref{eq:guinier} to the Guinier region is mainly valid up to \( qR_g<1.3 \). In this restricted \(q\)-range, too few data points are available for a reliable data analysis. Therefore, an extrapolation using the spherical form factor \( f_{sph}(q,R) \) over the range available before the first minimum has been employed instead to obtain \(R_g\) and \(I(0)\).

\begin{figure}
\caption{Experimental squared radius of gyration  as a function of the solvent electron density. Equation \eqref{eq:gyration} is fitted to the data and shown as a thick line. The vertical and horizontal asymptotes correspond to \( \rho_0 \) and \( \tilde R^{\,2}_{g,c} \) respectively.}
\scalebox{0.45}{\includegraphics{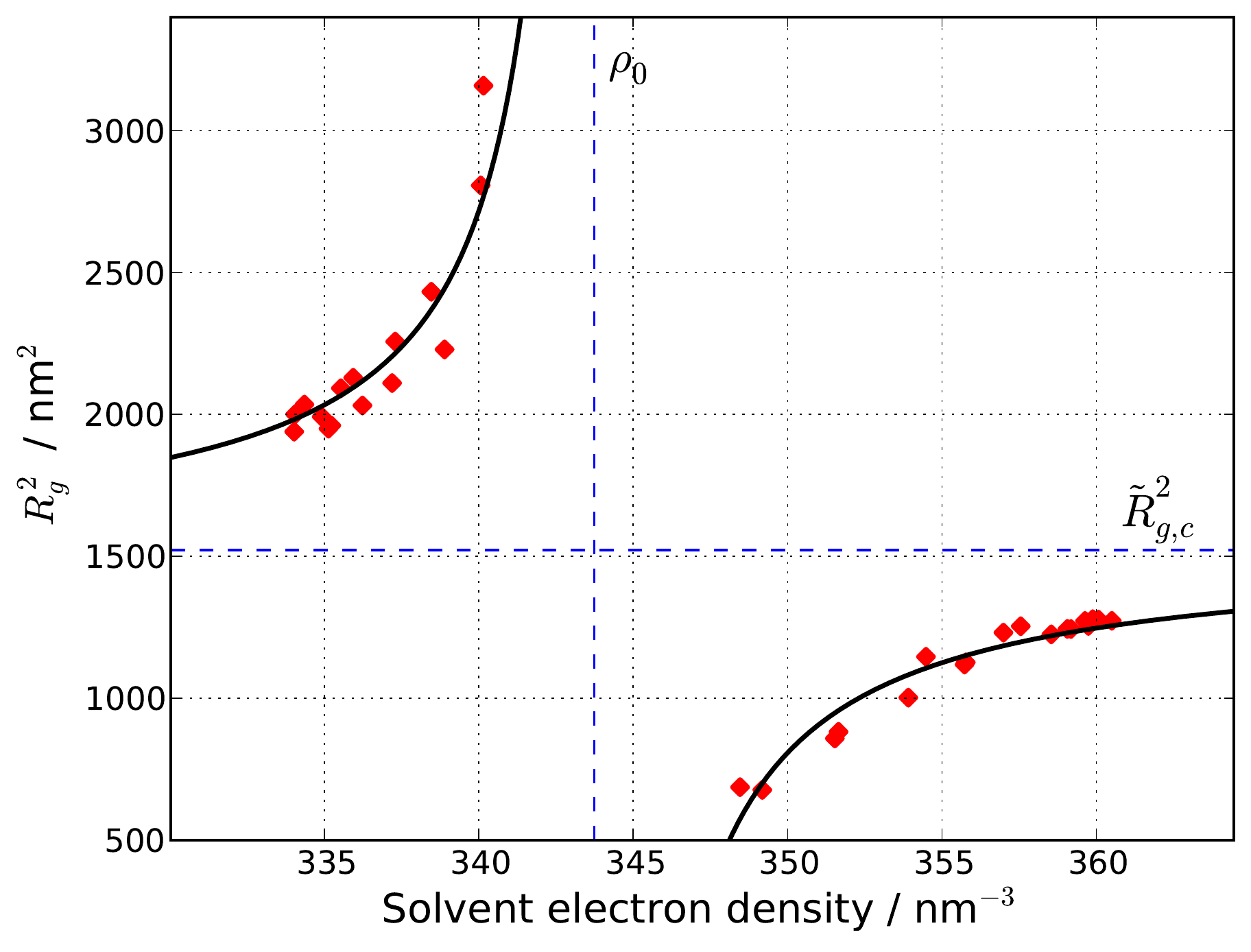}}
\label{fig:5}
\end{figure}

As described in \(\S\)\ref{sec:guinier}, the radius of gyration of a heterogeneous particle in a contrast variation experiment should behave according to equation \eqref{eq:gyration}. In figure \ref{fig:5}, the experimental squared radius of gyration is displayed as a function of the suspending medium electron density. The best fit to the measured data with values \(\rho_0=343.7\) nm\(^{-3}\), \( \tilde R_{g,c}=39.0\) nm, \(\tilde \alpha=4470\) nm\(^{-1}\) and \(\tilde\beta=0\) nm\(^{-4}\) is shown by the solid line. 

The positive value of \(\tilde\alpha\) validates the hypothesis that a more dense polymer (probably PMMA) surrounds a lighter core (PS) \cite{Stuhrmann2008}. The calculated average electron density of the particle \(\rho_0\) suggests a very thin layer of PMMA shell around the PS core, due to the proximity of its value to the polystyrene electron density (339.7 nm\(^{-3}\) ). The value of \( \tilde\beta=0\) proves a concentric model, where core and shell share the same centre. Using the same polydispersity value of \(22.8\,\%\) obtained in the fitting process, the value for the particle shape radius of gyration results in \(R_{g,c}=36.9\) nm and the external radius of the particle can be calculated assuming the particle as a spherical object (\( R_g^2=\frac{3}{5}R^2 \)). This calculation gives \( R=47.6\) nm, which is only 2.1 nm smaller than the calculated external radius \(R=49.7\) nm, though it might be underestimated due to the choice of a possibly inflated polydispersity. 

\begin{figure}
\caption{Experimental zero-angle intensity as a function of the solvent electron density. The function corresponding to equation \eqref{eq:I0} is fitted to the data and shown as a thick line. The minimum in the parabola corresponds to \( \rho_{solv}=\rho_0 \).}
\scalebox{0.45}{\includegraphics{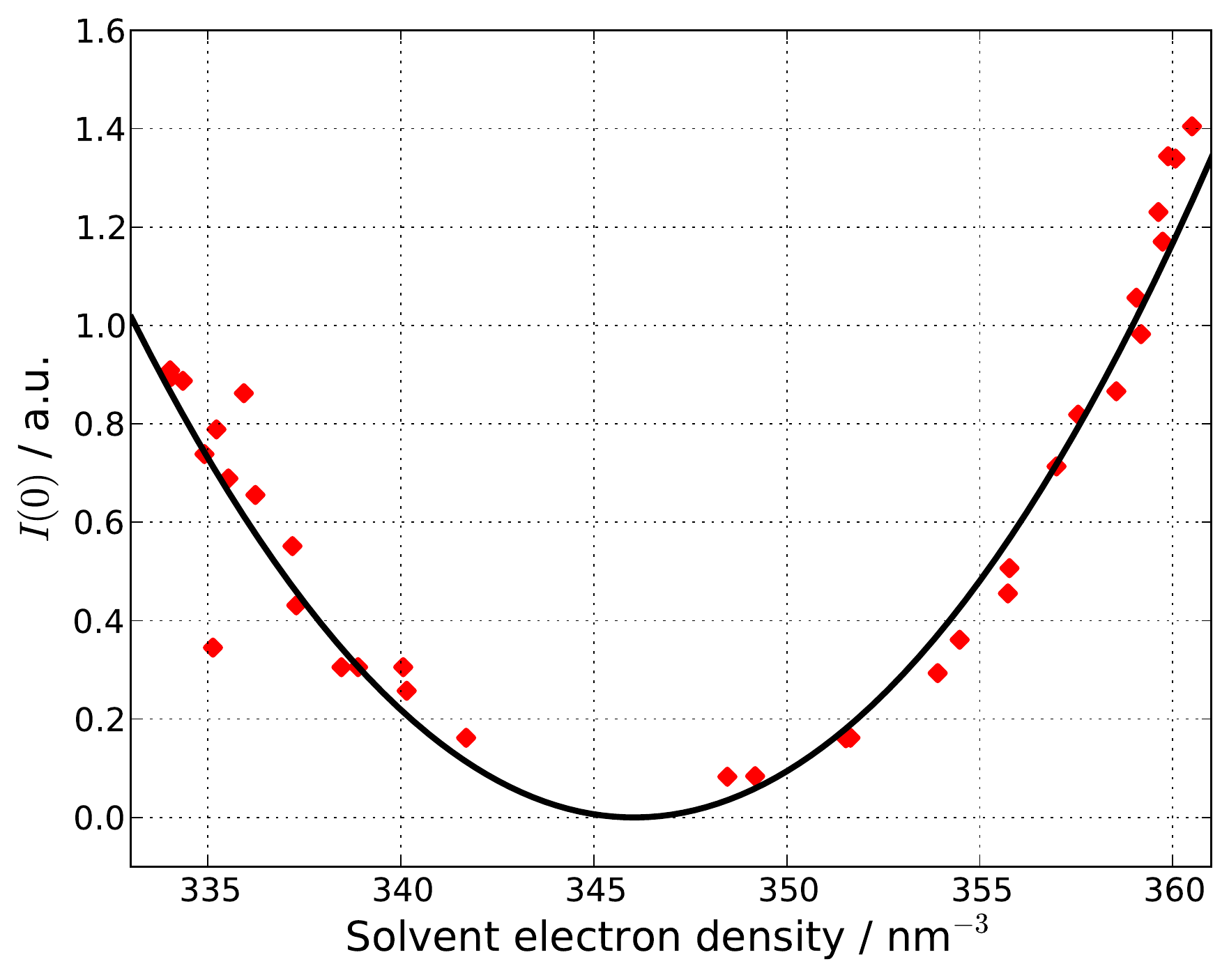}}
\label{fig:6}
\end{figure}

Using the same set of 40 scattering curves, the behaviour of the zero-angle intensity under the contrast variation is also investigated by fitting equation \eqref{eq:I0} to the experimental \(I(0)\), as depicted in figure \ref{fig:6}. A minimum in the curve is observed at \(\rho_{solv}=346.0\) nm\(^{-3}\), which corresponds to the value of the average electron density of the particle. This value is in very good agreement with the result obtained by fitting the core-shell form factor. It is also noticeable that the minimum intensity is approximately 0, which means that the effective average density of the ensemble is equal to the average density of the particle \cite{Avdeev2007}. This result further legitimates the previously made assumption that the ratio between the particle components' volumes is constant independent of the polydispersity and hence \(  \tilde \rho_0 = \rho_0  \).

\subsection{Consistency of the results}
Table \ref{tab:composition} summarizes the results of all three presented methods. From the first two isoscattering points, values for the external radius and an upper bound to the polydispersity degree have been derived. Focusing on the Guinier region of the scattering curves, a value for the average electron density of the particles is found using the radius of gyration as well as the zero-angle intensity, the values of which differ by 2.3 nm\(^{-3}\). By fitting a core-shell model, an external radius of \(R=49.7\) nm and an average electron density \(\rho_0=345.9\) nm\(^{-3}\) have been obtained, which are in considerable agreement with the previous results, i.e., the values determined by the other methods are included in their confidence ranges, except for the \(\rho_0\) calculated with the radii of gyration.

\begin{table}
\caption{Comparison of the different methods presented in this article to evaluate SAXS contrast variation data.}
\begin{tabular}{l|ccc}
 & \( R\) (nm)    & \(\rho_0\) (nm\(^{-3}\)) & \(p_d\) (\(\%\))\\
\hline
 Core-shell fitting &  49.7\(\pm 2.8\)   &     345.9\(\pm 1.5\)      & 22.8\(\pm 6.0\) \\
 Isoscattering point* &  50.5 &     -          & $<24$  \\
 Radius of gyration &  47.6**      &     343.7      & -    \\
 Zero-angle intensity &  -    &     346.0      & -    \\ \hline
\end{tabular}\\[0.8\baselineskip]

\begin{minipage}{7.5cm}
	\begin{raggedright}
	*Mean value of \(q^{\star}_1\) and \(q^{\star}_2\) \\
	\setlength{\parindent}{-25pt}
	**Using the $p_d$ from the core-shell model fitting
	\end{raggedright}
\end{minipage}
\label{tab:composition}
\end{table}

From these results, it is evident that the radius of gyration interpretation produces the most deviant values. This might be founded in the complicated function fitted to the data and the reduced availability of $q$-range employed to obtain \( R_g \).

The resulting polydispersity degree of the measured particles from the model fit is in agreement with the upper limit obtained with the radii of gyration. Nevertheless the polydispersity is the parameter determined with the largest uncertainty in the fitting process and therefore this result must be considered with care.

It can be concluded that the different approaches show consistent and complementary results about the size distribution of nanoparticles with radial inner structure, especially for the external radius of the particle and its average electron density. A precise value for the polydispersity degree could not be obtained as explained previously, although a credible upper limit to the polydispersity degree of $24\,\%$ could be given.

\section{Conclusions}
\label{sec:conclusion}
This article demonstrates that it is possible to perform continuous contrast variation for light nanoparticles by means of a density gradient and to collect a large quantity of SAXS curves, which can be analyzed with complementary approaches to reveal a consistent insight into the size distribution and the inner structure of the suspended nanoparticles. 

By means of a model-free analysis of the experimental data based on the isoscattering point theory, an average particle diameter of 101 nm was obtained. The analysis of the Guinier region of the scattering curves shows that the radial inner structure of the particles consists of a thin, more dense layer coating the polystyrene core. Complementing these results, a core-shell model fit showed that the core component of the particle had exactly the same electron density expected for polystyrene and the shell was composed of a compound with a density below that of PMMA. This core-shell structure was expected for chemical reasons due to the different hydrophobicity of PS compared to MMA and MAA.

Considering the similar electronic composition of these polymers and the average electron density of the particle $\rho_0=346$ nm\(^{-3}\), an average physical density of the particles of $\rho=1.07$ g/cm\(^{3}\) can be calculated. The precision in the determination of this density proves this technique as a useful tool and an alternative to other techniques like isopycnic centrifugation \cite{Vauthier1999,Arnold2006,Sun2009}, widely used with biomacromolecules.

Nevertheless, future applications of this technique must consider the limited density range accessible with an aqueous sucrose solution, which restricts the applicability to light particles. More dense solutions prepared with heavy salts could be used as an alternative, but they might compromise the stability of the particles and lead to more complicated handling of the sample due to a decreased diffusion timescale. Other possible methods that vary the contrast of a single medium have already been proposed (e.g. ASAXS \cite{Stuhrmann2007}), although a system fulfilling the requirements must be found and a large complementary dataset might be difficult to acquire.

\ack{The authors are grateful to Dr A. Hoell of the Helmholtz-Zentrum Berlin for the research cooperation with the HZB SAXS instrument, to Dr Z. Varga of the Hungarian Academy of Science for fruitful discussions and valuable comments, to L. Cibik and S. Langner of PTB for their technical support and to Dr. N. Buske of MagneticFluids for providing us with Galden\textregistered PFPE SV90. This work was funded by the European Metrology Research Programme (EMRP). The EMRP is jointly funded by the EMRP participating countries within EURAMET and the European Union.}

\bibliography{iucr}

\end{document}